
\magnification=\magstep1
\baselineskip= 24 true pt
\vsize=22.5 true cm
\hsize=16.5 true cm
\centerline {\bf Realisations of $GL_{p,q}(2)$ quantum group
  and its coloured extension }
\centerline {\bf  through a novel Hopf algebra with five generators }
\vskip  1.25 true cm
\centerline  { B. Basu-Mallick\footnote*{e-mail address:
biru@imsc.ernet.in}}
\centerline  {\it The Institute of Mathematical Sciences }
\centerline {\it  C I T Campus, Taramani, Madras-600113, India }
\vskip 2  true cm
\noindent {\bf Running Title :}   A new Hopf algebra
related to $GL_{p,q}(2)$  quantum group
\vskip .23 true cm
\noindent {\bf Abstract}

A novel Hopf algebra   $ ( {\tilde G}_{r,s} )$,  depending  on two
deformation parameters and five generators, has been constructed.
This  $ {\tilde G}_{r,s}$   Hopf algebra might be considered
as some quantisation  of classical  $GL(2) \otimes GL(1) $ group, which
contains the standard $GL_q(2)$ quantum group (with $ q=r^{-1} $)
as a Hopf subalgebra. However,  we interestingly observe that
the two parameter deformed $GL_{p,q}(2)$ quantum group can also be realised
through the generators of this $ {\tilde G}_{r,s}$   algebra,
provided the sets of deformation parameters $p,~q$ and $r,~s$ are related
to each other in a particular fashion.
Subsequently we construct the invariant noncommutative planes associated
with  $ {\tilde G}_{r,s}$ algebra and show how the two well known Manin planes
corresponding to   $GL_{p,q}(2)$ quantum group can  easily be reproduced
through such construction. Finally we consider the `coloured' extension
of $GL_{p,q}(2)$ quantum group  as well as corresponding Manin planes
and  explore their intimate connection with the `coloured' extension of
 $ {\tilde G}_{r,s}$   Hopf  structure.
\hfil \eject
\noindent {\bf 1. Introduction }

Quantum groups and related algebras [1-4], which have originated
 from the study of integrable models,  are
 now finding their applications in diverse branches of
physics and mathematics [5-8]. Since the underlying symmetry of many
physical systems are governed by the quantum group structures,
it is expected that their  representations would
also  play a significant role in determining the behaviour of such systems.
This is one of the prime reasons why a lot of work  has been done
in recent years for building up  the representations of quantised universal
enveloping algebras [6,9].  However, it may be noted that
the representations of their dual objects,
i.e. quantum groups, have  not  received that much
 attention   and yet  under rapid  developement [10-14].
Still lesser amount of   progress has been made   for the case of
 multiparameter quantum groups [15], which are   more difficult
 to handle  due to the presence of multiple  deformation parameters
 in the commutation relations. So it is natural to hope  that a better
 understanding on some common origin of single and multiparameter
quantum groups would be helpful for  building up
 their  representation theory.

Moreover,  as has been found recently [16], it is possible
to construct  a `colour' parameter dependent  quantum group which interestingly
reproduces  the  standard $GL_q(2)$ and $GL_{p,q}(2)$  quantum groups as its
subalgebras, at the monochromatic   limit of  colour parameters.
The  invariant Manin planes corresponding to this
infinite dimensional  quantum group would  also
depend explicitly on the colour parameters. Consequently this coloured
quantum group might be looked as some further extension  of the
two parameter deformed $GL_{p,q}(2)$  case, and one  may expect that  similar
generalisations  can be done  for other multiparameter quantum groups.
 However, though having many nice properties,  such  coloured   algebras
look rather cumbersome in comparison with their standard counterparts.
So for  obtaining    a deeper insight  about these   algebras
and  constructing  their  representations, it is  useful  to enquire
whether they can be  connected  to  some other algebras    with
relatively  simpler structure.

The purpose of the present article is to shed some light
on the above issues, by focussing on the simplest  $GL_q(2)$ and
$GL_{p,q}(2)$ quantum groups as well as their  coloured extension.
In our investigation we are able to find
a novel Hopf algebra ( denoted by ${\tilde G}_{r,s}$ ) depending on
two deformation parameters and five generators.
 Curiously, the first four generators of such Hopf algebra form
a Hopf subalgebra, which coincides exactly with the single
parameter dependent $GL_{q}(2)$ quantum group when $q=r^{-1}$. However,
 it turns out that, that the  two parameter deformed $GL_{p,q}(2)$
quantum group can  also be realised through
the generators of this  ${\tilde G}_{r,s}$  Hopf algebra, provided
the  sets of deformation parameters $p,~q$ and $r,~s$ are related
to each other in  a particular fashion. So this new
algebra  with five generators is found to be connected in a strange  way
to both $GL_q(2)$ and $GL_{p,q}(2)$ quantum groups.
In sec.2 we introduce this  Hopf structure and explore
its relation with $GL_{p,q}(2)$ quantum group.

 Generators of  ${\tilde G}_{r,s}$   Hopf
algebra can interestingly be arranged in a  ($3 \times 3 $ ) matrix form,
which satisfies the quantum Yang-Baxter equation for a certain  choice
of braid group representation.  So this Hopf algebra
might also be considered  as a
quantum group, and one may naturally  ask  whether there exists any
invariant  noncommutative  plane related to such structure.
In sec.3 we investigate on this problem and  able to
find  such quantum planes  possessing some curious
properties. For example, in contrast to the  case of a usual  quantum group
where one gets   only  two  Manin planes, in the present  case
 we obtain   as many as eleven such  invariant noncommutative
planes. Moreover the algebra  of corresponding
coordinates contain a free parameter, which cannot be determined from
the   associated  ${\tilde G}_{r,s}$  structure.  We also discuss in sec.3 how
some composite functions of these noncommuting coordinates are  able
to reproduce the two well known Manin planes related to the $GL_{p,q}(2)$
quantum group. Finally in sec.4 we consider the  colour parameter dependent
 extensions  of $GL_{p,q}(2)$ and ${\tilde G}_{r,s}$ quantum groups,
 and then   discuss  their close connection to each other.
 Sec. 5 is the concluding section.
\hfil \break
\vfil \eject
\noindent { \bf 2. Realisation of   $GL_{p,q}(2)$ quantum group
through  a novel Hopf algebra }
\vskip .1 true cm
Before coming to the main results  of this section let us
summarise briefly some  basic properties of the  $GL_{p,q}(2)$ quantum group.
This  quantum group might be  generated by four  elements $a,~b,~c$ and $d$
satisfying the relations [17,18]
  $$
   \eqalign {
   ab~=~p~  ba~,~~ac~=~q~  ca~,
   ~~&db ~=~  q^{-1}  bd~ ,~~dc ~=~ p^{-1}   cd~, \cr
      bc ~=~{q\over p } cb~,~~~[a,d] ~=
  &~ ( p - q^{-1} ) ~ bc ~,   }
\eqno (2.1)
 $$
where the deformation parameters $p$ and $q$ are some nonzero complex
numbers. For the particular case $p=q$,  (2.1) reduces to the algebra of
  one parameter deformed $GL_{q}(2)$ quantum group.
As it is well known, quantum groups can in general be  constructed
  from the solutions
of quantum Yang-Baxter equation (QYBE)   given by [3]
  $$           R~T_1~T_2 ~~=~~ T_2~T_1~R    ~,   \eqno (2.2)        $$
where $R$ is a nonsingular $(N^2 \times N^2 )$-dimensional matrix with usual
$c$-number valued elements,
 $T_1 = T \otimes {\bf 1} , ~ T_2 =  {\bf 1} \otimes T ~$  and
noncommuting elements of  $T$-matrix are identified with the
generators of a quantum group.  Due to the associativity of QYBE (2.2),
$R$-matrix   satisfies  the spectral parameterless Yang-Baxter
equation (YBE) given by
$$  R_{12} ~R_{13}~R_{23}~=~R_{23}~R_{13}~R_{12}~,
\eqno  (2.3) $$
where $R_{ij}$ acts on the direct product of three vector spaces and is
nontrivial only  on the $i$th and $j$th spaces
 ( e.g. $R_{12} = R \otimes {\bf 1} $ etc. ).
All  commutation relations in (2.1), corresponding to
the  $GL_{p,q}(2)$ quantum group, can easily  be generated from
QYBE  (2.2) by taking  the   $T$-matrix in the form [18]
  $$  T = \pmatrix { {a} & {b} \cr {c} & {d}  } ,     \eqno (2.4) $$
and  the  related  $R$-matrix satisfying YBE (2.3) as
$$
   R ~=~\sqrt {p \over q }~ \pmatrix {   {  p^{-1} } & {} & {} & {} \cr
     {} & {1} &  { p^{-1} - q } & {} \cr
   {} & {0} & { q p^{-1}  } & {} \cr {} & {} & {}
& {  p^{-1}    }  } ~. \eqno (2.5)
$$
The inverse of $T$-matrix   (2.4), defined through the relation
$~TT^{-1} ~=~T^{-1}T ~=~{\bf 1 } $,  may be given by
$$
    T^{-1}  = \delta^{-1}
\pmatrix { {d} & {- q^{-1}b} \cr {-qc} & {a}  } ,     \eqno (2.6)
$$
where $\delta^{-1} \delta = \delta \delta^{-1} = 1 $, and $\delta $
is the quantum determinant:
$$  \delta ~=~ ad ~-~ q~cb~.   \eqno (2.7)  $$
However, in contrast to the case of one parameter deformed $GL_q(2)$ group,
 the above quantum determinant is no longer a central element of $GL_{p,q}(2)$
 algebra and by using (2.1) it is easy to see that $\delta $ obeys the
 commutation relations
$$
a\delta ~=~ \delta a ~,~~b \delta ~=~ q p^{-1}  ~\delta b ~,~~
c \delta ~=~  q^{-1} p ~ \delta c ~,~~d\delta ~=~ \delta d ~. \eqno (2.8)
$$
The coproduct ($\Delta $) for  $GL_{p,q}(2)$ Hopf algebra may   be obtained
by substituting  the    form of $T$-matrix (2.4) in  the   expression
$~ \Delta T = T { \buildrel  . \over \otimes  }T ,~$
where the symbol  ${ \buildrel  . \over \otimes }$
signifies ordinary matrix multiplication  with tensor multiplication
of algebra.  The other Hopf  structures, like
 co-unit ($\epsilon$ ) and antipode ($ K $ )
for    quantum groups,  might  in general  be expressed  as
$$\epsilon (T_{ij}) = \delta_{ij} ~, ~~ K (T) = T^{-1} \eqno (2.9)  $$
and for   the particular  case of
 $GL_{p,q}(2)$ quantum group they  may be   explicitly given by
$$
\eqalign  {
\epsilon (a) ~=~ \epsilon (d) ~=~ 1, ~~
&\epsilon (b) ~=~ \epsilon (c) ~=~ 0,  \cr
K(a) ~=~\delta^{-1} d , ~~  K(b) ~=~ -  q^{-1}~  \delta^{-1} b, ~~
&  K(c) ~=~ -  q ~\delta^{-1} c, ~~K(d) =  \delta^{-1} a  ~. }
\eqno (2.10)
$$
Here it is assumed  that the inverse of  quantum determinant,
  satisfying the relations
$~ \Delta ( \delta^{-1} ) = \delta^{-1} \otimes \delta^{-1}, ~
\epsilon  ( \delta^{-1} ) = 1 , ~ K( \delta^{-1} )  = \delta $,
is included in the algebra.

Now for constructing  a new  realisation of this $GL_{p,q}(2)$ quantum group,
 we consider in the following another Hopf algebra  $ {\tilde G}_{r,s}$
    containing  five generators $A,~B,~C,~D $  and $F$.
 These generators satisfy the algebraic relations
  $$
   \eqalign  {
   AB~=~r^{-1}~BA  ~,~~AC~=~  r^{-1}~CA ~,
   ~~&DB ~=~  r~BD ~,~~DC ~=~r~CD ~, \cr
      BC ~=~ CB~,~~~[ A,&D ]~= ( r^{-1} - r ) BC ~, }
   \eqno (2.11a)
  $$
  $$
AF~= ~ FA,~~ BF~=~s^{-1}~FB~,~~ CF~=~ s ~FC~, ~~ DF~=~ FD~,
\eqno (2.11b)
$$
where the  two deformation parameters $r$ and $s$
 are arbitrary nonzero complex numbers. Notice that the elements
 $A,~B,~C$ and $D $   of  $ {\tilde G}_{r,s}$ algebra
form a subalgebra, which obeys the commutation relations (2.11a) and
 exactly  coincides with the   single parameter deformed
$GL_q(2)$  algebra  when  $~ q=r^{-1} $. It is easy to see that,
again quite similar to the $GL_q(2)$ case,
the Casimir operator of  algebra (2.11) would be given by
$$
{\cal D} ~=~ AD~-~ r^{-1} ~BC ~ .  \eqno (2.12)
$$
The coproduct for this   $ {\tilde G}_{r,s}$  structure may be written as
$$
\eqalign {
\Delta (A) ~=~  A \otimes A ~+~  B \otimes C ~,~~
&\Delta (B)  ~=~  A \otimes B  ~+~  B \otimes D ~,~~\cr
\Delta (C)  ~=~  C \otimes A  ~+~  D \otimes C ~,~~
\Delta (D)  ~=~  &C \otimes B  ~+~  D \otimes D ~,~~
\Delta (F)  ~=~  F \otimes F~.   }
\eqno (2.13)
$$
Finally, other Hopf relations like co-unit and antipode
for   $ {\tilde G}_{r,s}$  will take the form
$$
\eqalign  {
\epsilon (A) ~=~ \epsilon (D) ~=~ \epsilon (F) ~=~1, ~~
&\epsilon (B) ~=~ \epsilon (C) ~=~ 0,  \cr
K(A) ={ \cal D }^{-1} D , ~  K(B) = - r { \cal D }^{-1} B, ~
  K(C) = -  r^{-1} &{\cal D }^{-1} C, ~K(D)
=  {\cal D }^{-1} A  ,~K(F) = F^{-1} , }
\eqno (2.14)
$$
where  $ { \cal D }^{-1} $
is  included in the algebra and it   satisfies   the relations like
$ \Delta ( { \cal D }^{-1} ) $= $ { \cal D }^{-1} \otimes $ $
{ \cal D }^{-1}~,$
$ \epsilon  (  { \cal D }^{-1} )  = 1 , ~$
$K ( { \cal D }^{-1} )~=~ { \cal D } .~$
Now  by using the
 expressions (2.11)-(2.14), one can easily  define  a consistent
Hopf structure through the  polynomials made of
$ {\tilde G}_{r,s}$ generators, which would  satisfy all axioms [1,19]
of Hopf algebra:
$$
\eqalignno{
m ( id \otimes m ) ~=~ m ( m \otimes id ) ~,~~
 ( id \otimes & \Delta ) \Delta ~=~( \Delta \otimes id ) \Delta ~, ~~~~~~~~
&(2.15a,b) \cr
( id \otimes \epsilon ) \Delta ~=~ ( \epsilon \otimes id ) \Delta
{}~=~id ~,~~ m ( id \otimes  K  ) \Delta &
{}~=~ m ( K \otimes id ) \Delta ~=~ {\bf 1} \cdot \epsilon ~,~~~~~
& (2.15c,d) \cr
\Delta ( xy ) ~=~ \Delta (x) \Delta (y)~,~~ \epsilon (xy) ~=~
\epsilon (x) \epsilon (y)&  ~,~~ K(xy) ~=~ K(y)K(x)~,~~~~~
& (2.15e,f,g)  }
$$
where $m$ denotes  the multiplication operation ( $ m (x\otimes y)
= xy $ ) and $id $ is the  identity transformation.
 Moreover, the elements
$A,~B,~C$ and $D$ of $ {\tilde G}_{r,s}$
evidently  form a Hopf subalgebra, which coincides with
the $GL_q(2)$ quantum group.

Next we  attempt  to explore  whether there exists any connection between
$GL_{p,q}(2)$ quantum group and  the newly defined
 $ {\tilde G}_{r,s}$    Hopf algebra,
both of which  contain two deformation parameters.
For this  purpose we  propose a simple realisation of
   $GL_{p,q}(2)$ generators    through  the
 elements  of  $ {\tilde G}_{r,s}$ algebra as
$$
a ~=~ F^N A ~,~~ b ~=~ F^N B~,~~  c ~=~ F^N C ~,~~ d ~=~ F^N D~,
\eqno (2.16a)
$$
where $N$ is any  fixed nonzero  integer. Notice that the above
  relations  can also
be written in a convenient matrix form given by
  $$   \pmatrix { {a} & {b} \cr {c} & {d}  } ~=~  F^N~
   \pmatrix { {A} & {B} \cr {C} & {D}  } ~.  \eqno (2.16b)
  $$
Now by using  the   $ {\tilde G}_{r,s}$ algebra
  (2.11a,b) it is not difficult to verify
that (2.16) indeed gives us a realisation of $GL_{p,q}(2)$ algebra
(2.1), provided the two sets of deformation parameters  $(p,q)$ and $(r,s)$
are related through the equations
$$   p ~=~ r^{-1}~ s^N~, ~~~ q ~=~ r^{-1}~ s^{-N}~ .  \eqno (2.17) $$
Observe that when $N \neq 0 $, one can easily
invert the above equations  to find out the values of deformation parameters
$(r,s)$ for any given  value of $(p,q)$ and consequently
 realisation like (2.16) is always possible.
Moreover by using    (2.16) and  the coproduct  for
$ {\tilde G}_{r,s}$ algebra (2.13) along with  its  homeomorphism property
(2.15e), one can easily recover the standard coproduct for $GL_{p,q}(2)$
quantum group.  Finally, it is also possible to get back the co-unit
and antipode   for  $GL_{p,q}(2)$ (2.10), with the help
of relations (2.14),  (2.15f,g) and (2.16). Thus  we see  that
the full Hopf algebra structure related to $GL_{p,q}(2)$ quantum
 group can in fact be reproduced
through the realisation (2.16).

Though we shall not  consider  here about  the representations of
  $GL_{p,q}(2)$ quantum group,
let us make some remarks on how the realisation
like (2.16) might be  useful in this context.  As we have already observed,
in contrast to $GL_{p,q}(2)$ algebra (2.1)  where
  parameters $p$ and $q$ both appear in  the commutation relations,
 a large subalgebra of ${\tilde G}_{r,s}$ depends on a single
deformation parameter.  So it  might be relatively easier
to construct first the representations of ${\tilde G}_{r,s}$ algebra and
then use the realisation (2.16) for obtaining the representations of
$GL_{p,q}(2)$ quantum group.
Moreover, it is interesting to notice that
the mapping (2.17) from $(r,s)$-plane to $(p,q)$-plane depends on the choice
of integer $N$. So by taking different values of $N$,
a single point on the $(r,s)$-plane  can be mapped
over   infinite number of discrete points, which
would satisfy the conditions  like
$$
p_N q_N = r^{-2} ~,~~ { p_{N+1} \over p_N } ~=~s~,~~
 { q_{N+1} \over q_N } ~=~ s^{-1} ~,
\eqno (2.18)
$$
where $(p_N,q_N)$ denotes  a point on the  $(p,q)$-plane
  corresponding  to  the  $N$-th mapping.
As evident from eqn. (2.18), all these discrete points will lie on a
hyperbola and their density in a given region would be controlled by the
parameter $s$. Therefore we find   that from the representation of
 ${\tilde G}_{r,s}$   algebra  for a particular value of deformation
parameters $(r,s)$, one can build up the representations of $GL_{p,q}(2)$
quantum group for infinitely many discrete values of corresponding
deforming parameters. Thus there exists a rather  interesting
structural connection  among the  $GL_{p,q}(2)$ representations for these
discrete values of $(p,q)$  parameters. Furthermore from eqn. (2.18) we
   see that by  taking the limit $s\rightarrow 1$,
 for any fixed value of $r$, one can increase
the density of such discrete  points  arbitrarily.  So  the  properties
of  ${\tilde G}_{r,s}$  algebra  at the neighbourhood of $s=1$ line may have
some  particular significance in the context of $GL_{p,q}(2)$ representations.
 In the next section we
shall explore further the close connection  beween ${\tilde G}_{r,s}$
and $GL_{p,q}(2)$ structures, by studying   the corresponding
noncommutative  planes.
\vskip .5 true cm
\noindent {\bf 3. Quantum plane for ${\tilde G}_{r,s}$   algebra  }
\vskip .1 true cm
For constructing invariant quantum planes  [20-21]
 associated with the
 ${\tilde G}_{r,s}$ Hopf algebra, we first try to cast this algebra
in the form of a quantum group and define a $(3 \times 3)$ $T$-matrix
through the corresponding generators as
$$
T ~=~ \pmatrix {  {A} & {B} & {0} \cr {C} & {D} & {0} \cr {0}& {0} & {F}   }~.
\eqno (3.1)
$$
Now  it is easy to verify that,  for the above form
$T$-matrix and the choice of ($9\times 9$) $R$-matrix given by
$$
R ~=~r~ \sum_i    ~e_{ii} \otimes e_{ii} ~+
 ~\sum_{i \neq j} ~f_{ij} ~\cdot ~e_{ii} \otimes e_{jj} ~+~(r-r^{-1})
 \sum_{ i<j  } ~e_{ij} \otimes e_{ji}  ~,
\eqno (3.2)
$$
where $i,~j ~\in ~ [1,3],$
 $f_{12}=f_{23}=1,~f_{13} = s $ and $ f_{ij} = f_{ji}^{-1}  $,
 QYBE (2.2) is able to reproduce all  relations
  of ${\tilde G}_{r,s}$   algebra (2.11).
Moreover the $R$-matrix (3.2)  turns out to be a solution of
 the spectral parameterless YBE (2.3).
So we find that the associative
 ${\tilde G}_{r,s}$   algebra  can in fact be written
in the form of a quantum group. Furthermore,  by taking
the $T$-matrix as (3.1) and using the expression of  coproduct for quantum
groups:
$~ \Delta T = T { \buildrel  . \over \otimes  }T ,~$ one can easily
recover the coproduct relations (2.13)
corresponding to  ${\tilde G}_{r,s}$   algebra.
Finally, by constructing  the inverse of $T$-matrix (3.1) as
$$
T^{-1} ~=~ \pmatrix {   { {\cal D}^{-1}D } & { - r {\cal D}^{-1}B }
&  {0}  \cr
{ - r^{-1}  {\cal D}^{-1}C  }  &      {\cal D}^{-1}A  &  {0}   \cr
{0} & {0} & { F^{-1} }  }
$$
and using the  expression (2.9), one can also get back the related
counit and antipode structures (2.14).

It is worth noting  at this point that the most general Hopf algebra
 which can be generated from the $R$-matrix (3.2),
 by assuming  all elements of  corresponding $(3\times 3)$   $T$-matrix
to be nonzero objects,
is a multiparameter dependent   extension of $GL_q(3)$ quantum  group.
Then putting by hand the additional
restriction on such general $T$-matrix that some of its elements
are equal to  zero, one can reproduce the   ${\tilde G}_{r,s}$  structure
 and so this Hopf algebra with five generators might    be interpreted
  as a quotient of multiparameter deformed
 $GL_q(3)$ quantum  group. However it is important to notice that the elements
of general $T$-matrix, which have been put to  zero in the above mentioned
case,
  actually form a biideal of multiparameter deformed
$GL_q(3)$  quantum  group and due to  this crucial reason  one gets
  here a consistent Hopf structure
through the quotient procedure. A   different kind   of
   quotients,  for some  other  quantum groups,   have  been
studied very recently in the context of braided Heisenberg algebra [22].

There exists another interesting way to look at the
 ${\tilde G}_{r,s}$  structure, by considering it a two parameter dependent
quantisation of classical  $~GL(2) \otimes GL(1)~$ group. Evidently,
the first four generators of ${\tilde G}_{r,s}$, i.e.
$A,~B,~C,~D$ correspond to $GL(2)$  group at the classical level and
 the remaining generator $F$ is related to $GL(1)$ group. Moreover, while
the parameter $r$ arises from  the deformation of $GL(2)$ group,
the other parameter $s$ curiously enters through
 the nontrivial  cross commutation relations
between $GL_r(2)$ and $GL(1)$ group elements at the quantum level. Similar
type
of quantisations,  for the  particular case of some semisimple Lie groups,
has been considered  in   ref.23 and subsequently used to construct the
 deformation of    wellknown  Lorentz algebra.

Now to find out the invariant
Manin  planes associated with  ${\tilde G}_{r,s}$
quantum group, we  take a vector ${\vec X}~=~( X_1,~X_2,~X_3 ) $
having three noncommuting components and  define the action
of  $T$-matrix (3.1)  on this vector as
$$
{\vec X}'~~=~~ T ~{\vec X} ~;~~~ \pmatrix { {X_1'} \cr {X_2'} \cr {X_3'} }
{}~~=~~
 \pmatrix {  {A} & {B} & {0} \cr {C} & {D} & {0} \cr {0}& {0} & {F}   }~
 \pmatrix { {X_1} \cr {X_2} \cr {X_3} }   ~. \eqno (3.3)
$$
Then,  as usual, we assume the elements of  $T$-matrix to be commuting
with the components of  vector ${\vec X}$ and by using ${\tilde G}_{r,s}$
 algebra (2.11) try to find out the
form of commutation relations among the components  $X_i$,
which would  be preserved under the transformation (3.3).
 Through  explicit calculation
we find that there exist eleven   such sets of commutation relations and two
among these sets   may be  given by
$$
X_1 X_2 ~=~  r^{-1} ~ X_2 X_1 ~,~~X_2X_3 ~= ~ k~X_3X_2 ~,~~X_1X_3 ~=~
k  s^{-1}~  X_3 X_1 ~,  \eqno (3.4a)
$$
and
$$
X_1^2 ~=~ X_2^2 ~=~ 0 ~,~~
X_1 X_2 ~=~ - r~ X_2 X_1 ~,~~X_2X_3 ~= ~ k~X_3X_2 ~,~~X_1X_3 ~=~
ks^{-1}  X_3 X_1 ~, \eqno (3.4b)
$$
respectively,
where $k$ is an arbitrary number. Thus the relations (3.4a)
and (3.4b) provide  us  two Manin planes related to the   ${\tilde G}_{r,s}$
quantum group. To obtain other Manin planes one may  notice
that  each of the  four  sets of relations  given by
$$
\eqalign {
&1) ~~X_3^2 ~=~0~,~~~~~~~2)~~
X_2X_3 ~= ~ k~X_3X_2 ~,~~X_1X_3 ~=~ ks^{-1}~  X_3 X_1 ~,  \cr
&3) ~~
X_1 X_2 ~=~  r^{-1} ~ X_2 X_1 ~,~~~~~~
4) ~~X_1^2 ~=~ X_2^2 ~=~ 0 ~,~~
X_1 X_2 ~=~ - r~ X_2 X_1 ~,}
$$
 would also  remain invariant under the action of $T$-matrix (3.1). However,
it is interesting to observe  further that,  no  smaller subset of
any of these  four set of relations
 is able to generate independently  some other
 invariant  noncommutative  plane.
 Consequently, the four noncommutative  planes
 related to these sets  are `irreducible' in nature. Now by taking advantage
of  the block diagonal form of $T$-matrix (3.1), it is not difficult
 to prove  that  all possible  consistent  combinations of these  four
sets of relations, would also give us new invariant quantum planes.
  For example,
 by combining the sets 2) and 3) one can easily reproduce the earlier relation
(3.4a). By taking various combinations of the sets 1) to 4),
  we can  similarly  generate   all of the
eleven Manin planes related to ${\tilde G}_{r,s}$  structure. Evidently,
in any of such  combinations the mutually inconsistent
 sets 3) and 4) would not appear simultaneously.
So this  procedure of deriving composite Manin planes from the four fundamental
blocks, is very similar to the construction of
 reducible representations of group
theory by taking direct sum of its  irreducible representations.
It may also be noted that the noncommuatative
planes related to the   usual  quantum groups  do not depend on  any extra
parameter, apart from those which are already present in the corresponding
algebra. But in the set of commutation relations  (3.4a,b)
 and in some other  similar
 sets, we interestingly find the existence of free parameter $k$, which
cannot be fixed through our quantum group relations (2.11).

After finding out the  noncommutative  planes related to
${\tilde G}_{r,s}$ structure, we like to     explore
in the following  their connection with the
 Manin planes  associated with
  $GL_{p,q}(2)$ quantum group.  For this purpose we denote the
 two dimensional quantum plane related to the latter
case as $~{\vec x}~=~ (x_1 , ~ x_2 ) $, on which   the
 $T$-matrix (2.4) would   act like
$$
\pmatrix { { x_1' } \cr {x_2'} } ~~=~~ \pmatrix {  {a} & {b} \cr {c} & {d}  }~
\pmatrix { {x_1} \cr {x_2}  }  ~. \eqno (3.5)
$$
Next  we  propose a simple  realisation of  this two dimensional
 ${\vec x}$ plane  through
previously discussed three dimensional ${\vec X}$ plane
associated with  the ${\tilde G}_{r,s}$ quantum group:
$$
\pmatrix { {x_1} \cr {x_2}  } ~~=~~ X_3^N ~ \pmatrix {  {X_1} \cr {X_2} }~,
\eqno (3.6)
$$
where as before $N$ is a  fixed nonzero integer.
 Substituting now the expressions (3.6)
and (2.16b) in the r.h.s. of (3.5) and using  subsequently
 the relation (3.3) we get
$$
\pmatrix { { x_1' } \cr {x_2'} } ~=~ F^N  \pmatrix
{  {A} & {B} \cr {C} & {D} } \cdot
X_3^N  \pmatrix {  {X_1} \cr {X_2} }~=~
(FX_3)^N \pmatrix {  {A} & {B} \cr {C} & {D} }
 \pmatrix {  {X_1} \cr {X_2} }~=~
X_3'^N ~ \pmatrix {  {X_1'} \cr {X_2'} }~.   \eqno (3.7)
$$
Comparing now  the  eqns. (3.6) and (3.7)  one may  interestingly
observe  that, the transformed vectors ${\vec x'}$
 and  ${\vec X' }$  are related to each other  in exactly  same way
as  their original  counterparts  ${\vec x}$ and ${\vec X}$.
 Now for  ${\tilde G}_{r,s}$  invariant quantum  planes,
  the form of commutation relations
among the components of ${\vec X'}$ must be identical with  that of ${\vec X}$.
So from the  structural similarity of  eqns. (3.6) and (3.7)
 it follows that, the  form of commutation relations among the components
of vector ${\vec x'}$ will  also be identical with that of
  vector ${\vec x}$. Consequently,
  by using the realisation (3.6)  and the commutation
relations corresponding to a ${\tilde G}_{r,s}$ invariant plane,
  if  we are able to derive some
bilinear   relations for the components of  vector ${\vec x}$,  then
 that would  automatically provide us an invariant Manin plane related
to $GL_{p,q}(2)$ quantum group.
For example,
by using  (3.4a) and (3.4b) respectively, along with (3.6),
we will get two sets of commutation relations:
$$
 x_1x_2 ~=~ r^{-1}s^{-N}~x_2x_1~;~~~~~~~x_1^2=x_2^2 = 0~ ,~~x_1x_2 ~=~
-{ 1\over r^{-1} s^N }~ x_2x_1 ~. \eqno (3.8a,b)
$$
Notice that the above relations are free from the extra
parameter $k$, though it was present in the original commutation
relations (3.4a,b). Moreover,  by using (2.17), one can easily  rewrite
(3.8a,b) in terms of familiar $p,~q$ variables and   this
    would exactly  reproduce
  the two well known Manin planes [18] related to  $GL_{p,q}(2)$
 quantum group. It may also be noted that if one starts with
  other invariant   quantum  planes corresponding  to ${\tilde G}_{r,s}$
 structure, then by using corresponding commutation relations
and realisation (3.6) it would not be possible to derive  some nontrivial
bilinear  relation for the components of vector ${\vec x}$. So
these cases do not lead us to   any new   $GL_{p,q}(2)$  invariant plane.

Thus from the above analysis  we find  an interesting
alternative method for constructing the noncommutative  planes associated with
 $GL_{p,q}(2)$ quantum group,
through that of the ${\tilde G}_{r,s}$ quantum  group. In the next
section our aim is  to  discuss how a simple modification of this
 ${\tilde G}_{r,s}$ structure also provides  a very convenient and
natural basis,  for realising  the
`coloured' extension of $GL_{p,q}(2)$  quantum group.
\vskip .5 true cm
\noindent {\bf 4. Coloured extension of $GL_{p,q}(2)$ and ${\tilde G}_{r,s}$
 quantum groups }
\vskip .1 true cm
There exists an intriguing possibility of extending
the standard  quantum group
relations,  by parametrising the corresponding generators through
some continuous `colour' variables and redefining the associated
algebra, co-product etc. in such a way that all Hopf algebra properties
would still remain  preserved. In  analogy with the standard cases,
the algebra for  these
coloured quantum groups might be constructed by slightly modifying
the form of QYBE (2.2) as  [3,16]
     $$   R^{(\lambda , \mu )} ~T_1(\lambda )~T_2(\mu ) ~~=~~
     T_2(\mu )~T_1(\lambda ) ~R^{(\lambda ,\mu )}    ~,   \eqno (4.1)   $$
where $\lambda ,~ \mu $ are the colour parameters.
Notice that the usual $T$-matrix in (2.2) has been replaced by $T(\lambda )$
or $T(\mu )$ in the above expression and the same $ij$-th element of
$T(\lambda )$ and $T(\mu )$ matrices, i.e. $T_{ij}(\lambda ) $ and
$T_{ij}(\mu )$, would correspond to different generators when $\lambda \neq
\mu $.
 So this type of   algebra practically
depends on infinite number of generators, in contrast to finite dimensional
cases encountered in the previous sections.
However due to the structure
of QYBE (4.1), generators of only two different colours can at most
appear simultaneously in the algebraic relations. Moreover the
subalgebra   formed by the elements of any  particular colour
should coincide with some usual quantum group related  algebra,
since at the monochromatic limit  $\lambda = \mu $
(4.1) effectively reduces to the more common form of QYBE (2.2).
The coproduct of algebra (4.1) may be given by the simple relation
$$
 \Delta T(\lambda )
{}~~=~~ T(\lambda )~ { \buildrel  . \over \otimes  } ~T(\lambda ) ,~\eqno (4.2)
$$
where the symbol  ${ \buildrel  . \over \otimes }$
signifies  ordinary matrix multiplication  with tensor multiplication
of algebra.

Due to the associativity of QYBE (4.1),   $R^{ (\lambda , \mu ) }$
satisfies the YBE like
$$
R_{12}^{ +(\lambda , \mu ) }~ R_{13}^{ +(\lambda ,\gamma ) }~
R_{23}^{ +(\mu ,\gamma ) }~ =~
                            R_{23}^{ +(\mu ,\gamma ) }
     ~R_{13}^{+ (\lambda ,\gamma ) } ~R_{12}^{ +(\lambda , \mu ) },
\eqno (4.3)
$$
which  naturally reduces to eqn. (2.3)
at the limit $\lambda = \mu = \gamma $. We shall further assume
that the  $R^{ (\lambda , \mu ) }$-matrix satisfying (4.3) would be
upper or lower triangular in form, in analogy with the structure of
$R$-matrices related to  the standard quantum groups. This type of
 $R^{ (\lambda , \mu ) }$-matices are also closely connected with the
coloured braid group representations, which have attracted much attentions
in recent years [24-28].

 A special form of $(4 \times 4)$
  $R^{ (\lambda , \mu ) }$-matrix,
which can  be derived by taking  the fundamental representation
of universal ${\cal R}$-matrix associated with the $U_q(gl(2))$ quantised
algebra [26], may be given by
     $$
R^{ (\lambda , \mu ) } ~=~ \pmatrix {
{ t^{1- (\lambda - \mu ) }    } & {} & {} & {} \cr
{ \eqalign { \cr  } }
&  { t^{ \lambda + \mu  } }  &  {  (t-t^{-1} )     }
& {} \cr
{} & {0} & {  t^{- ( \lambda + \mu ) }  }  &  {} \cr
{} & {} & {} & { t^{1 + (\lambda - \mu ) } }    } ~.
\eqno (4.4)
$$
The above  solution of YBE (4.3)
is particularly interesting,  since at the
monochromatic limit $ \lambda = \mu = \Lambda $  it reduces exactly
 to the  $R$-matrix  (2.5)  associated with  $GL_{p,q}(2)$
quantum group, when   the   parameters $p$ and $q$  are given through
the relations
$$
p ~=~ t^{ -1 + 2 \Lambda } ,~~~
q~=~ t^{ - ( 1 +  2 \Lambda ) } ~. \eqno (4.5)
$$
So it  is   expected  that the solution (4.4) would lead to
 a `coloured'  quantum group, whose generators
corresponding to any  particular colour will reproduce
the $GL_{p,q}(2)$ algebra as a subalgebra.
Such coloured extension of $GL_{p,q}(2)$ quantum group may be
  constructed by taking   the corresponding
        $ (2\times 2)$  $T(\lambda )$-matrix as [16]
  $$
    T(\lambda ) =
  \pmatrix { {a(\lambda )} & {b(\lambda )} \cr {c(\lambda )}
  & {d(\lambda )}  } ,     \eqno (4.6)
  $$
     and  inserting this $T(\lambda )$  along with  $R^{(\lambda , \mu )}$
(4.4) in the QYBE (4.1). By expressing then  QYBE in  elementwise form
we would   get  a large number of independent
  algebraic relations, which may  be  grouped  together
in two different sets. The first set is formed through the relations
$$
    \eqalignno {
    a(\lambda )b(\mu ) ~=~ t^{-1+2\lambda }~b(\mu ) a(\lambda ) ~,~~
  & a(\lambda )c(\mu ) ~=~ t^{-1-2\lambda }~c(\mu ) a(\lambda ) ~,~~& (4.7a,b)
  \cr
    d(\lambda )b(\mu ) ~=~ t^{1+2\lambda }~b(\mu ) d(\lambda ) ~,~~
&    d(\lambda )c(\mu ) ~=~ t^{1-2\lambda }~c(\mu ) d(\lambda ) ~,~~&(4.7c,d)
   \cr
    b(\lambda )c(\mu ) = t^{-2(\lambda +\mu ) }~
    c(\mu ) b(\lambda ) ,~~[ a(\lambda )&, d(\mu ) ] =
    - (t-t^{-1})t^{ \lambda +\mu  }~b(\mu ) c(\lambda ),
    & {} \cr &~ &(4.7e,f )    }
$$
and the second set contains the remaining independent relations given by
  $$
  \eqalignno {
  a(\lambda )b(\mu ) ~=~t^{\lambda - \mu } ~a(\mu ) b(\lambda ) ~,~~
    &a(\lambda )c(\mu ) ~=~t^{-\lambda + \mu } ~a(\mu ) c(\lambda ) ~,
    ~~& (4.8a,b)
  \cr
    d(\lambda )b(\mu ) ~=~ t^{\lambda - \mu }~d(\mu ) b(\lambda ) ~,~~
    &d(\lambda )c(\mu ) ~=~t^{-\lambda + \mu } ~d(\mu ) c(\lambda ) ~,
{}~~&(4.8c,d)
   \cr
    b(\lambda )c(\mu ) ~=~
  b(\mu ) c(\lambda ) ~,~~ a(\lambda ) d(\mu ) ~=~a(\mu )& d(\lambda ) ~,~~
  a(\lambda ) a(\mu ) ~=~  a(\mu ) a(\lambda ),
    & (4.8e,f,g )    \cr
  b(\lambda ) b(\mu ) ~=~ t^{2(\lambda - \mu )}  b(\mu ) b(\lambda ) ~,~~
 c(\lambda ) c(\mu ) ~=~ &t^{-2(\lambda - \mu )} c(\mu ) c(\lambda )~,~~
d(\lambda ) d(\mu ) ~=~ d(\mu ) d(\lambda ) ~. & {}\cr
&~ &(4.8h,i,j) }
$$
So the above two sets   define explicitly  the algebra of
the present  coloured quantum group.
Now  for  finding  out the commutation relations among the
elements of a particular colour, one have to  take the  $\lambda =
\mu = \Lambda $ limit for both of  the sets (4.7) and (4.8). By taking this
limit for the set (4.7) we find that it would exactly reproduce
the $GL_{p,q}(2)$ algebra (2.1), when the values of $p,~q$  parameters
are given through the expression (4.5).
 On the other hand, all relations in the second set (4.8) would
become trivial at this $\lambda = \mu $ limit. These results
demonstrate that the
 $GL_{p,q}(2)$ algebra can be reproduced as a subalgebra,
 from the monochromatic limit  of our coloured algebra.
 Moreover the relation (4.5) shows that, by selecting  diffrent
values of the monochromatic limit $\Lambda $, one can also change
the values of deformation parameters $p$ and $q$. So this coloured algebra
actually contains infinite number of such
 $GL_{p,q}(2)$  subalgebras, with continuously varying values of
corresponding deformation parameters.

 The coproduct for this extended algebra
might  be obtained by simply substituting the $(2 \times 2 )$
$T(\lambda )$-matrix
(4.6) in the general expression (4.2). One can also define the quantum
determinant corresponding to the $T(\lambda )$-matrix (4.6) as
$$
  \delta (\lambda ) ~=~ a(\lambda ) d(\lambda ) ~-~t^{-(1 + 2 \lambda) }~
  c(\lambda )  b(\lambda ) ~.
  \eqno (4.9)
  $$
By using now (4.7) and (4.8) it is not difficult to verify that
the above quantum determinant satisfies the relations like
  $$
  \eqalignno {
  a(\lambda ) \delta (\mu ) = \delta (\mu ) a(\lambda ),~
&b(\lambda ) \delta (\mu ) = t^{-4 \mu }~ \delta (\mu ) b(\lambda ),~
c(\lambda ) \delta (\mu ) = t^{4 \mu }~ \delta (\mu ) c(\lambda ),~~~~~~~~~
&(4.10a,b,c)  \cr
d(\lambda )& \delta (\mu ) ~=~ \delta (\mu ) d(\lambda )~,~~~
\delta (\lambda ) \delta (\mu ) ~=~ \delta (\mu ) \delta (\lambda )~.
 ~~~~~~~~~~~~~~~&(4.10d,e)    }
$$
{}From the above expression
 we see that $\delta (\mu )$ does  not  commute
 with all elements of $T(\lambda )$ unless $\mu = 0 $, and so  only
  $\delta (0)$ would  be a Casimir operator of full coloured algebra.
 However
the relation (4.10e) reveals  the interesting fact that the  determinants
associated with different colours form a set of mutually commuting
operators.   By assuming now  that
     the   inverse of $\delta (\lambda )$
             exists  for all values of $\lambda $, one can
  write down the inverse of the $T(\lambda )$-matrix
(4.6) as
$$
T(\lambda )^{-1}  ~~=~~\delta (\lambda )^{-1}     \pmatrix {
{d(\lambda ) } & { -t^{1+2\lambda } b(\lambda ) }  \cr
{ - t^{-1-2\lambda } c(\lambda )  } & { a(\lambda ) }   } .
$$
    Finally the co-unit   and antipode  relations
    for this  coloured case may be given by
  $$
    \eqalign {
      &\epsilon \left ( a(\lambda ) \right ) ~=~
      \epsilon \left ( d(\lambda ) \right ) ~=~ 1~,~~~
      \epsilon \left ( b(\lambda ) \right ) ~=~
      \epsilon \left ( c(\lambda ) \right ) ~=~0~, \cr
&k \left ( a(\lambda ) \right ) ~=~ \delta (\lambda )^{-1} d(\lambda ) ~,~~~
k \left (b(\lambda )\right )~=~- t^{1+2\lambda }~
\delta (\lambda )^{-1}  b(\lambda ) ~,
 \cr
 &k \left (c(\lambda )\right )~=~
- t^{-1-2\lambda }~\delta (\lambda )^{-1}  c(\lambda ) ~,~~~
k \left ( d(\lambda ) \right ) ~=~\delta (\lambda )^{-1 }
 a(\lambda ) ~,}\eqno (4.11)
  $$
\vskip .15 true cm
\noindent   with $ \Delta  \left ( \delta (\lambda )^{-1}  \right )
= \delta (\lambda )^{-1}  \otimes  \delta (\lambda )^{-1},$
  $~ \epsilon \left ( \delta (\lambda )^{-1}  \right ) =1 , $
 $~ k \left ( \delta  (\lambda )^{-1}  \right ) = \delta (\lambda )  , $
 and it is easy to check that they
 would  satisfy all conditions   of a   Hopf algebra.

One of the most interesting features of the above described coloured
extension of $GL_{p,q}(2)$ quantum group is the existence of
 corresponding
noncommutative   planes [16]. Such quantum planes may be constructed
by inserting   the colour parameter to the original two dimensional
 $GL_{p,q}(2)$ quantum plane: $ {\vec x }(\lambda ) ~=~ (~ x_1(\lambda ),~
x_2(\lambda )~ )$. So the coordinates of  this infinite dimensional
quantum  plane not only  depend on a discrete set of indices,
but also on a continuously variable colour parameter. With the help
of $T(\lambda )$-matrix (4.6),
one may   define  the transformation on these coloured coordinates as
$$
 \pmatrix { {x_1'(\lambda ) } \cr {x_2'( \lambda )}  } ~~=~~
 \pmatrix { {a(\lambda )} & {b(\lambda )}
 \cr {c(\lambda )} & {d(\lambda ) }  }
 \pmatrix { {x_1(\lambda ) } \cr {x_2( \lambda )}  }  .
              \eqno (4.12)
      $$
Next, in analogy with the standard cases, we assume that
the matrix elements of $T(\lambda )$  would commute with
 $x_i(\mu )$ for all vaules of $\lambda, ~\mu $,
and using the  algebra (4.7), (4.8)
try to find out the commutation relations between the coordinates
of different colours which will be  invariant under
the above  transformation.  Being guided by the form of QYBE (4.1),
we also make the simplifying assumption
 that the coordinates of  only  two different  colours    can
appear simultaneously in these commutation relations. Now
 by somewhat  lengthy but  straightforward
  calculation      we    find  that there exist two
such sets of invariant  relations given by
     $$   \eqalign {
             x_1(\lambda ) x_1(\mu ) ~=~t^{\lambda - \mu } ~
                 x_1(\mu )  x_1(\lambda )~,~~
            & x_1(\lambda ) x_2(\mu ) ~=~ t^{-(1+\lambda + \mu )} ~
                 x_2(\mu )  x_1(\lambda )~,    \cr
        x_1(\lambda ) x_2(\mu ) ~=~
                 x_1(\mu )  x_2(\lambda )~,~~ &
         x_2(\lambda ) x_2(\mu ) ~=~t^ { -\lambda + \mu } ~
                 x_2(\mu )  x_2(\lambda )~,   }  \eqno (4.13)
    $$
    and
   $$  \eqalign {
      x_1(\lambda ) x_1(\mu ) ~= ~&  x_2(\lambda ) x_2(\mu ) ~=~0~,~~\cr
       x_1(\lambda ) x_2(\mu ) ~=~-~ t^{1-\lambda - \mu } ~
                 x_2(\mu )  x_1(\lambda )~, & ~~
        x_1(\lambda ) x_2(\mu ) ~=~
                 x_1(\mu )  x_2(\lambda )~.  }
    \eqno (4.14)
    $$
It is worth noting  that at the monochromatic
limit $\lambda = \mu  $, (4.13) and (4.14) recover the expressions
 (3.8a)  and (3.8b) respectively, after
a trivial  transformation of the related  parameters. So
the two  coloured quantum planes, defined through the
relations (4.13) and (4.14), contain the standard
    $q$-plane  and  its exterior plane corresponding to
       $GL_{p,q}(2)$ quantum group as their  subspaces.

In spite of the fact that the algebra defined through the
relations (4.7) and (4.8) exhibit many interesting
properties, it is  dependent on the colour parameters in a quite
complicated way. So for having some  deeper insight into  this kind of
algebraic structure,  it is helpful to investigate whether it can be realised
through some other algebra of  comparatively simpler form. For this purpose
we  propose a  simple  coloured extension
of  ${\tilde G }_{r,s}$ Hopf algebra discussed in sec.2,
by keeping its  first four generators $A,~B,~C,~D$  to be
 independent of the colour parameters,
 but  replacing the fifth generator
$F$  by the continuous set of generators $F(s)$:
  $$
   \eqalign  {
   AB~=~r^{-1}~BA  ~,~~AC~=~  r^{-1}~CA ~,&
   ~~DB ~=~  r~BD ~,~~DC ~=~r~CD ~, \cr
      BC ~=~ CB~,~~[ A,D ]~= ( &r^{-1} - r ) BC ~,
 ~~ [ F(s), F(s') ] ~=~ 0, \cr
AF(s)~=~ F(s)A~,~~ BF(s)~=~s^{-1}F(s)B~,~~
 &CF(s) ~=~ s~ F(s)C~,~ ~ DF(s)= F(s)D,~ }
\eqno (4.15)
$$
where the colour parameters
$s$ and $s'$ may take arbitrary values.  One can easily check that
the associativity property holds for the  above algebra. Moreover,
it is possible to define
the corresponding  coproduct, antipode etc.,
 in close analogy with the   ${\tilde G }_{r,s}$ case,
which would obey all axioms of a Hopf algebra. Genetators of the
  algebra  (4.15)  can  also be cast
in the form of a  coloured quantum group satisfying
QYBE (4.1),  when the parameters $\lambda , ~\mu $  are replaced
by  the present   colour parameters $s,~s'$.
 The $T(s )$-matrix   might be  defined for  this case
  by simply  substituting  the element $F$
with  $F(s)$ in the expression  (3.1) and the  associated
$R^{(s,s')}$-matrix would  be a coloured generalisation of the $R$-matrix
(3.2), which  we are not giving  here explicitly.
{}From the structure of  $T(s)$-matrix it is evident that,  the
   noncommuting plane corresponding to this
 coloured quantum group would be composed of
the components $~X_1,~X_2 ,~ X_3(s) $ and these components
 will be transformed like
$$
X_1'~=~ AX_1 + BX_2~,~~X_2' ~=~ CX_1 + DX_2 ~,~~ X_3(s)'~=~ F(s) X_3(s) .
\eqno (4.16)
$$
 Now by using the algebra (4.15) one can   find  that,
 just like the case of  ${\tilde G }_{r,s}$ quantum group, there
exists    eleven    invariant noncommutative  planes in the
present case
and two among them   may be   given through the commutation relations
$$
X_1 X_2 =  r^{-1}  X_2 X_1 ,~X_2 X_3(s) =  k X_3(s) X_2 ,
{}~X_1 X_3(s) =   k  s^{-1}  X_3(s)  X_1 , ~[X_3(s), X_3(s')] = 0
\eqno (4.17a)
$$
and
$$
\eqalign {
X_1^2 ~=~ X_2^2 ~=~ 0~ ,~~
&X_1 X_2 ~=~ - r~ X_2 X_1 ,  \cr
X_2X_3(s) ~= ~ k~X_3 (s) X_2 ~,~~X_1X_3(s)& ~=~   ks^{-1}~  X_3(s) X_1 ~,~
 ~[X_3(s), X_3(s')] ~=~ 0 }  \eqno (4.17b)
$$
respectively, where  $k$ is an arbitrary number. Other
invariant quantum  planes can  also  be derived easily, by  following
a  procedure similar to   which has been  described  in sec.3.
However these quantum planes will not be relevant for our
 purpose  of making connection with the  coloured extension of $GL_{p,q}(2)$
quantum group and so we are not giving their explicit expression.

Thus from the above discussion we see that
 ${\tilde G }_{r,s}$ quantum group admits its coloured extension in
 a very natural way  and the  associtated  Manin planes  can
also be  obtained   by slightly modifying the relations of ${\tilde G }_{r,s}$
invariant planes. However, the situation
 is widely  different for the case of coloured
 $GL_{p,q}(2)$ group, since the corresponding algebra and
 Manin planes are much more complicated in form in comparison with
 their standard counterparts.  Nevertheless
 we interestingly find that, similar to  the
realisation (2.16) related  to the standard  cases,
one can  build up  a realisation of the coloured extension
of $GL_{p,q}(2)$ algebra ((4.7),(4.8))  through that of the
${\tilde G }_{r,s}$ algebra (4.15)  as
$$
\pmatrix {
{a(\lambda ) } & { b(\lambda )} \cr {c(\lambda )} & {d(\lambda )}  }
{}~~=~~ F(s) ~\pmatrix { {A} & { s^{- {1\over 2} } B } \cr
 { s^{1\over 2 } C }  &   {D}  }~,
\eqno (4.18)
$$
where  we have taken $~ t = r $ as the relation among fixed deformation
parameters and $~t^{2\lambda }=s ~$ as the relation among variable
 colour parameters  corresponding to
  these two algebras.  Note that it is also possible
to  get a  slightly more general form of
 above realisation, which would depend on  a fixed nonzero integer $N$
analogous to the expression (2.16).  Since the
 coloured ${\tilde G }_{r,s}$ algebra   is  much simpler
 in form in comparison with the coloured extension of
 $GL_{p,q}(2)$ algebra,  one may  expect that
   realisations  like  (4.18)  would be useful for constructing the
representations corresponding to the latter case.
 However, in the following,  we shall restrict
ourselves by just showing  how the realisation (4.18) gives us
a convenient  shortcut way of constructing the Manin planes
(4.13), (4.14) related to coloured $GL_{p,q}(2)$ algebra.
  So, in analogy with  the approach adapted
 in sec.3, we propose a relation between the noncommuting
 coordinates associated with  coloured $GL_{p,q}(2)$   and
  ${\tilde G }_{r,s}$ algebras  as
$$
\pmatrix { {x_1(\lambda )} \cr {x_2(\lambda )} } ~~=~~ X_3(s )~
 \pmatrix { { s^{- {1\over 4} } X_1 }  \cr { s^{1\over 4} X_2 }   } ~,
\eqno (4.19)
$$
where $~s=t^{2\lambda }$.
Substituting now the expressions (4.18) and (4.19) in the r.h.s. of
(4.12) and using then the transformation law  (4.16) we find
$$
\pmatrix { { x_1'(\lambda ) } \cr {x_2'(\lambda ) } } ~
{}~=~
{}~F(s) X_3(s)  ~
 \pmatrix { { s^{- {1\over 4} } (~ A X_1 + B X_2 ~) }
 \cr { s^{1\over 4} (~ C X_1 + D X_2~) }   } ~=~
X_3'(s) ~
 \pmatrix { { s^{- {1\over 4} } X_1' }  \cr { s^{1\over 4} X_2' }   } ~.
\eqno (4.20)
$$
Comparing   the  eqns. (4.19) and (4.20)  it may be  observed
 that the transformed vectors ${\vec x~{}'}(\lambda )$
 and  ${\vec X' }(s)$  are related to each other in
exactly same way as  their original   counterparts  ${\vec x}(\lambda )$
 and ${\vec X}(s)$. So we  may conclude that,
if the vector ${\vec X }(s)$  corresponds to some
invariant quantum  plane associated
with the coloured ${\tilde G }_{r,s}$ algebra, and by using
 the realisation (4.19)  we are able to derive some nontrivial bilinear
relations among the components of vectors ${\vec x}(\lambda )$
and ${\vec x}(\mu  ),$
then those binear relations would  automatically  define  an
 invariant Manin plane related to the
coloured extension of  $GL_{p,q}(2)$ quantum group.
By using now
the expressions  (4.17a)  and (4.17b) respectively,  along with the
  realisation (4.19), we  interestingly
get two such  sets of bilinear relations and these two sets
 exactly  coincide with  our previous expressions (4.13) and (4.14).
Thus we find  here
an elegent alternative way of constructing the Manin planes
related to the coloured extension of
$GL_{p,q}(2)$ quantum group, since in this method
 we do not have to directly deal with
the complicated algebraic relations (4.7) and (4.8).
\hfil \break
\vfil \eject
\noindent {\bf 5. Concluding remarks  }
\vskip .1 true cm
In this article we have attempted to find some common origin of
single and multiparameter deformed $GL_q(2)$ and
$GL_{p,q}(2)$ quantum groups. Such attempt led us to construct
a  novel Hopf algebra $ {\tilde G}_{r,s}$,  which
depends  on two deformation parameters and five generators.
This  $ {\tilde G}_{r,s}$   Hopf algebra might    be considered
as a  quantisation  of classical  $GL(2) \otimes GL(1) $ group,
 and contains
the standard $GL_q(2)$ quantum group ( with $ q=r^{-1} $ )
 as a Hopf subalgebra. On the other hand,  the two parameter
deformed $GL_{p,q}(2)$ quantum group can also   be realised through the
generators of this $ {\tilde G}_{r,s}$   algebra, provided
 the set of deformation
parameters $p,~q$ and $r,~s$ are related to each other in a particular
fashion. Thus we  interestingly   find that both of
these $GL_q(2)$ and $GL_{p,q}(2)$
quantum groups can be generated from the same
 $ {\tilde G}_{r,s}$ structure, by taking different combinations of
its generators.

Subsequently we  try to
construct the  Manin  planes related to
the  $ {\tilde G}_{r,s}$  quantum group.  In contrast to the case of a usual
quantum group  where one gets only  two Manin planes, in the present case
we find as many as eleven  such invariant  planes. The block
diagonal nature of  corresponding $T$-matrix seems to be  responsible
for the existence of so many invariant  noncommutative  planes.
By using some of these invariant planes,  along with the connection between
$GL_{p,q}(2)$  and $ {\tilde G}_{r,s}$ quantum groups, we also  find an
alternative way  of constructing the
wellknown  Manin planes related to $GL_{p,q}(2)$ case.

The realisation of $GL_{p,q}(2)$ algebra through the
$ {\tilde G}_{r,s}$  generators becomes very useful in the context
of corresponding `coloured' extensions. As we have discussed in
sec.4, it is possible to construct an infinite dimensional
Hopf algebra   by inserting colour parameters in the original  $GL_{p,q}(2)$
generators.
One can also find the `coloured' Manin   planes associated with
such  extension of $GL_{p,q}(2)$ quantum group.
Though  having many interesting properties,
 this coloured $GL_{p,q}(2)$ algebra
 is rather complicated in form  and to find out the corresponding Manin planes
one has to go through straightforward but lengthy calculations.
On the other hand,  it turns out     that there exists
  a quite  simple    coloured extension of our
  $ {\tilde G}_{r,s}$  algebra and corresponding invariant quantum planes.
Furthermore  we find  that, similar to the case of their
standard counterparts, it is possible to give a
 realisation of coloured $GL_{p,q}(2)$ algebra  through the generators of
coloured $ {\tilde G}_{r,s}$  algebra.  By exploiting such realisation
we also present an  elegent alternative method for constructing
 the Manin planes associated with  the coloured $GL_{p,q}(2)$ algebra.

It seems that the relations established in this article, among  some
simplest type of quantum groups,  would be applicable
 to more general   contexts. For example, one may start with
a  multiparameter deformed version of $GL_q(N)$ quantum group and
investigate whether it can  also  be realised through another   Hopf algebra,
which has some extra generators but possibly contains a core subalgebra with
comparatively  simpler structure.
This type of realisations of multiparameter
quantum groups might be useful for constructing their
representations. Moreover such realisations
  may lead to the coloured extensions of associated
 quantum groups and  Manin planes in a very
natural way.
\vskip .5 true cm
\noindent {\bf Acknowledgments }
\vskip .1 true cm
I like to thank Prof. Sunil Mukhi for his interesting suggestions
which initiated this work. I am indebted to Prof. R. Jagannathan for his
   encouragment and   valuable  help at some parts of  this work.
Finally I wish to thank Prof. R. Chakrabarti for some illuminating
discussions.

\vfil \eject

\noindent {\bf References }
\vskip .2 true cm
\item {[1]}
 Drinfeld V G 1986 {\it Quantum groups } ( ICM Proceedings, Berkeley ) 798
\item {[2]} Jimbo M 1985 {\it
  Lett. Math. Phys.  } {\bf 10} 63
\item {[3]}  Faddeev L D,  Reshetikhin N Yu and  Takhtajan L A 1990
in  {\it Yang-Baxter equation  in integrable systems, }
ed. M. Jimbo ( World Scientific )
Advanced series in mathematical physics, Vol {\bf 10}, p. 299
\item  {[4]}
Macfarlane  A  J  1989 {\it J. Phys. } {\bf A22 } 4581;
\item {} Biedenharn L C 1989 {\it J. Phys. } {\bf A22 } L873
\item {[5]}    Wadati M,   Deguchi T and  Akutsu Y 1989
 {\it Phys. Rep. }  {\bf 180}  247
\item {[6]}  Saleur H and   Zuber J B 1990 {\it  Integrable lattice models
and quantum groups, }  Saclay preprint, SPhT/90-071
\item {[7]} Witten E 1989 {\it Comm. Math. Phys. } {\bf 121} 351
\item {[8] }  Bernard D  and  Leclair A 1990 {\it
Nucl. Phys. } {\bf B340} 721;
\item {} Watamura S 1993 {\it Comm. Math. Phys. }  {\bf 158} 67
\item {} Kratz H 1993 {\it Phys. Lett. } {\bf B317 } 60
\item {[9]} Fuchs J 1992 {\it Affine Lie algebras and quantum groups }
( Cambridge monographs on mathematical physics, Cambridge University Press ),
See Chapter 4 and the literature survey  therein
\item {[10]} Floratos E G 1990 {\it Phys. Lett. } {\bf 252B} 97
\item {[11]}  Weyers J 1990 {\it Phys. Lett.}  {\bf 240B}  396
\item {[12]}
 Chakrabarti R and Jagannathan R  1991 {\it J.Phys. } {\bf A24 } 1709
\item {[13] } Fu H C and Ge M L 1992 {\it J. Phys. } {\bf A25} L389
\item {[14]} Karimipour V 1993 {\it Lett. Math. phys. } {\bf 28} 207
\item {[15]}  Chakrabarti R and Jagannathan R 1991 {\it
J. Phys. }  {\bf A24} 5683;
\item {} Karimipour V 1993 {\it J. Phys. } {\bf A26 }  6277
\item {[16]}
 Basu-Mallick B 1993 {\it Quantum group and Manin plane
related to a coloured braid group representation, }
ICTP Preprint, IC/93/219
\item {[17]}  Demidov E E,   Manin Y, Mukhin E E,  Zhdanovich D V 1990 {\it
Prog. Theor. Phys. Suppl. } {\bf 102} 203
\item {[18]}  Schirrmacher A,  Wess J and   Zumino B 1991 {\it
Z. Phys. }  {\bf C49} 317
\item {[19]} Aschieri P and  Castellani L 1993
   {\it  Int. Jour. Mod. Phys.  }  {\bf A8}  1667
\item {[20]}  Manin Y 1988 {\it Quantum groups and noncommutative
geometry, }  Montreal Univ. \hfil \break
preprint, CRM-1561
\item {[21]}
  Schwenk J 1991 in {\it Quantum groups} eds.
Curtright T, Fairlie D, Zachos C
 ( World Scientific ) p.53
\item {[22]} Baskerville W K and Majid S 1993 {\it J. Math. Phys. }
{\bf 34 } 3588
\item {[23]}  Truini P and Varadarajan V S 1991 {\it Lett. Math. Phys. }
{\bf 21 }  287
\item {[24]} Murakami J 1991 in {\it Proc.  Int. Conf. of Euler Mathematical
School on Quantum groups, 1990}  ( Lecture notes in Physics,
Springer Verlag  ) p.350
\item {[25]} Akutsu Y and  Deguchi T 1991 {\it
Phys. Rev. Lett.  } {\bf 67} 777
\item {[26]} Burdik C and  Hellinger P 1992
J.Phys. {\bf A25} L1023
\item {[27]}  Kundu A,  Basu-Mallick B 1992
{\it J.Phys.  } {\bf A25}  6307
\item {[28]} Ge M L and Xue K 1993 {\it J. Phys. } {\bf A26} 281

\end